\pdfoutput=1

\documentclass[aip,jcp, reprint]{revtex4-1}
\usepackage{graphicx}

\draft 

\begin{document}


\title{Thermodynamics of Surface Defects at the Aspirin/Water Interface} 



\author{Julian Schneider}
\author{Chen Zheng}
\author{Karsten Reuter}
\affiliation{Chair for Theoretical Chemistry and Catalysis Research Center, \\Technische Universit\"at M\"unchen, Lichtenbergstr. 4, D-85747 Garching, Germany}
\email{karsten.reuter@ch.tum.de}


\date{\today}

\begin{abstract}
We present a simulation scheme to calculate defect formation free energies at a molecular crystal/water interface based on force-field molecular dynamics (MD) simulations. To this end we adopt and modify existing approaches to calculate binding free energies of biological ligand/receptor complexes to be applicable to common surface defects, such as step edges and kink sites. We obtain statistically accurate and reliable free energy values for the aspirin/water interface, which can be applied to estimate the distribution of defects using well-established thermodynamic relations. As a show case we calculate the free energy upon dissolving molecules from kink sites at the interface. This free energy can be related to the solubility concentration and we obtain solubility values in excellent agreement with experimental results.
\end{abstract}

\pacs{}

\maketitle 

\section{Introduction}

The production of a huge variety of both organic and inorganic solid molecular materials is achieved via crystallization of the substance from solution \cite{Garside1985, Shekunov2000}. Consequently, this research field has received a lot of attention, resulting in the development of advanced models of the interfacial attachment processes to predict crystal growth features based on only a minimum {\em a priori} input \cite{Burton51, Snyder09, Lovette12}. These models rely to a large extent on the notion that growth proceeds via rotating spirals of step edges and that incorporation of growth units takes place primarily at kink defects along step edges \cite{Burton51}. In essence, this concept allows to express the displacement velocity of advancing step edges as a function of only the kink densities and the net flux of molecules into these sites.

Regarding the reverse process, the dissolution of crystalline molecular solids into a given solvent, theoretical modelling has not advanced at the same pace. Although there is a general interest in dissolution structures and mechanisms, such as for instance etch pits \cite{Lasaga01}, the primary focus is in fact much more directed to the underlying kinetics. Absolute dissolution rates are of paramount importance, in particular in pharmaceutical applications, where the active ingredients, typically presented in crystalline form, have to be dissolved in order to enable their absorption into the blood circuit \cite{Amidon95}. A slow dissolution behavior can affect the entire pharmacokinetics of the substance and thus impair the efficacy of an otherwise promising drug candidate.

Although most of the basic mechanistic concepts can in principal be transferred in a straightforward way from growth to dissolution \cite{Cabrera}, the predictive quality of these microkinetic models in terms of quantitative growth or dissolution rates is generally hampered due to the lack of sufficiently accurate microscopic input quantities, such as rate constants and defect energetics. The prevalent approach to estimate defect formation free energies by the sum of the bond energies, which have to be broken upon creation of defect structures, has proven effective for the prediction of relative shapes of various organic crystals grown from solution \cite{Kuvadia2011,Lovette12}. As these approaches are based on potential energy differences of rigid structures, which, if at all, include solvent effects solely in a simplified implicit model, they might not be accurate enough for a quantitatively precise prediction of defect formation free energies, as required in reliable dissolution rate predictions. These concerns extend to the neglect of finite-temperature and entropic effects, as well as the inability of the implicit solvation description to account for hydrogen bonds between solvent molecules and polar moieties, present even in predominantly hydrophobic molecules, such as aspirin.

In the present work we therefore aim at formulating a more accurate molecular dynamics (MD) simulation scheme to obtain defect formation free energies at the solid/liquid interface. We apply the devised simulation scheme to the aspirin(001)/water interface, representing a prototypical active pharmaceutical ingredient (API). We proceed by outlining the thermodynamic cycle along which we calculate free energy differences, and by explaining in detail the free energy simulation methodology. After presenting and discussing the acquired simulation results for free energies of creating step edges, kink sites and for dissolving molecules from the interface, we interpret the values in a thermodynamic context using well-established relations to predict 2D-nucleation barriers, kink site densities, and solubility concentrations.

\section{Simulation Details}
\label{sec:simulation_details}

The interactions of aspirin and water molecules are described by the generalized AMBER force field (GAFF)\cite{GAFF} in combination with the TIP3P water model\cite{jorgensen1983}. We have assessed the reliability of the force fields in describing the crystal/liquid interface by comparison against experimental properties and quantum mechanical calculations.
While the detailed comparison and analysis of different force fields will be published elsewhere \cite{Greiner14}, we summarize here that the GAFF/TIP3P combination yields a very good balance between accuracy in structural properties and in the energetic description of various binding energies between aspirin/aspirin and aspirin/water pairs. Importantly, the solution enthalpy, i.e. the enthalpy change upon transferring a single aspirin molecule from the crystal into solution, is calculated as 26.7 kJ/mol \cite{Greiner14} and thus corresponds almost exactly to the experimental value of 27.0 kJ/mol \cite{Edwards}.

Notwithstanding, we note that the reliable simulation of proton transfer reactions, in this particular case the deprotonation of the carboxylic acid moiety, is beyond the capabilities of the common classical force fields, and consequently we disregard such effects in our simulations. Owing to the fact that aspirin is a weak acid (pK$_a$ = 3.6)
\cite{Edwards50}, we expect a large fraction of dissolved molecules to become de-protonated in aqueous solution at neutral pH. Strictly speaking, the simulation model thus corresponds to an acidic environment. Experimental results have revealed though, that the solubility changes only slightly from around 0.025 to approximately 0.021 mol/l when changing the solution environment from neutral to acidic \cite{Edwards}. Similar findings hold for the dissolution rates.
Furthermore, most of the considered interfacial defect structures in this study do not include molecular arrangements in which the carboxylic acid moieties are exposed towards solution and not saturated via hydrogen bonds with an opposite aspirin molecule. Accordingly, we expect an insignificant likelihood for a spontaneous deprotonation of these structures.

All simulations are performed using the 4.6 version of the GROMACS package\cite{GROMACS}. The production simulations are carried out at constant volume and constant temperature. The temperature is controlled through a Langevin thermostat \cite{vanGunsteren88} with a relaxation time of 0.5\,ps and a target temperature of 300\,K. This tight coupling of the system to a heat bath becomes particularly benificial when switching off the interactions of a certain set of molecules with their surroundings and thus removing the coupling to their natural heat bath. Pressure coupling during equilibration runs is achieved by employing a Parrinello-Rahman barostat \cite{ParrinelloRahman}. All Lennard-Jones potentials, as well as the short-range part of the Coulomb interactions are truncated at a cutoff distance of 1.0 nm, while long-ranged electrostatic interactions are calculated via the particle-mesh-Ewald (pme) method \cite{PME}.

To provide an absolute reference frame, a single molecular layer, either at the bottom or at the center of the slab, is restrained via harmonic potentials to the respective atomic crystal positions. In order to maximize the distance between this tethered layer and the defect structures we use the central layer for defects at both slab surfaces (i.e. step defects) and the bottom layer for defects at the top interface only. In general, the slab thickness is chosen sufficiently large to achieve a separation between restrained layer and interface defects larger than the Lennard-Jones cutoff distance.  This setup thus efficiently mimics a perfect underlying crystal substrate and provides an absolute reference frame for temporary position restraints to be applied to selected sets of defect molecules during the free energy simulations.

Bias potentials on collective variables other than cartesian coordinates, such as the mean-square-displacement (MSD) from a reference conformation, are employed via the PLUMED plugin \cite{bonomi2009}. For comparison, we furthermore calculate potential energy differences between rigid crystal arrangements, similar to the prevalent crystal engineering approach \cite{Lovette12}. Solvent effects are therein accounted for through an implicit solvation model. Instead of using the built-in generalized Born model augmented with the hydrophobic solvent accessible surface area term (GBSA) \cite{Still90} of the GROMACS code, we rely the \textit{Conductor-like Screening Model with Integer Charge} (COSMIC) \cite{Gale07}, as implemented in the \textit{General Utility Lattice Program} (GULP) \cite{Gale03}, which is better suited to model solid/liquid interface solvation. Notwithstanding, as we intend to use the results solely for an order of magnitude comparison to the accurate defect formation free energy values obtained through our MD based approach, we refrain from an extensive optimization of all COSMIC parameter values involved and simply employ the standard values of Ref. \cite{Gale07} instead.

\section{Model Systems}

\begin{figure*}
  \includegraphics[width=0.78\textwidth,clip]{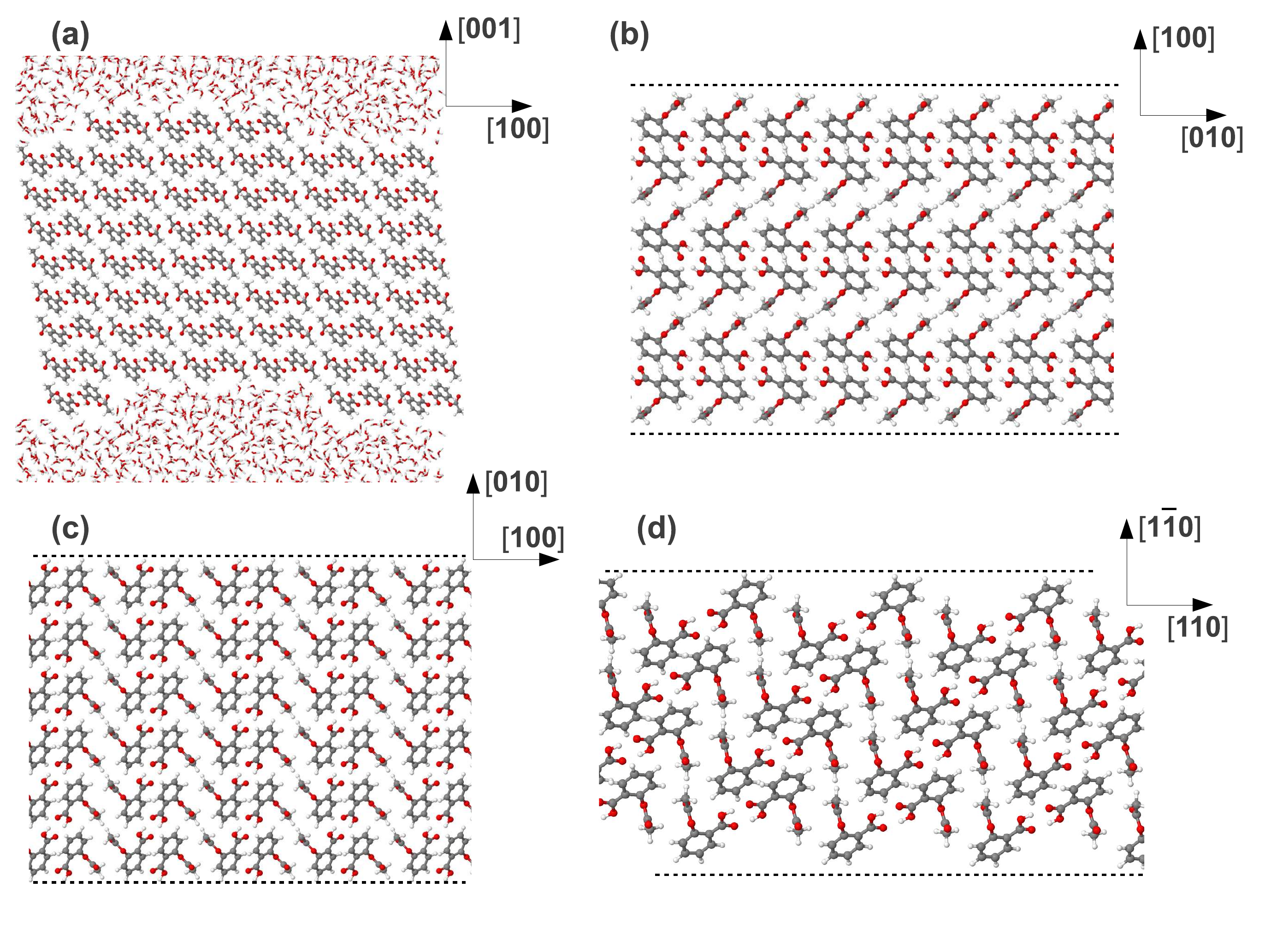}
  \caption{Step defect arrangements at the aspirin (001) surface. (a) Side view of the simulation supercell, illustrating the two equivalent steps arising at the edges of the protruding (indented) terrace at the top (bottom) of the aspirin slab. (b)-(d) Top views of the investigated step edges. For clarity only the terrace top layer between the two step edges at the top of the slab is displayed: (100) step edge (b), (010) step edge (c), and (1$\bar{1}$0) step edge (d). The respective step edges are marked by the dashed lines.}
  \label{fig:Step_edges}
\end{figure*}

The most common crystal structure of aspirin belongs to the $P2_1/c$ space group \cite{Kim85b}. Within this structure each unit cell contains two dimers of aspirin molecules, in which the two molecules of each dimer are held together by a pair of hydrogen bonds between their carboxylic acid groups. These bonds form the strongest intermolecular connections within the aspirin crystal.

In this study we exclusively consider defect structures at the aspirin (001)/water interface. The (001) facet forms one of the dominating surfaces of aspirin crystals grown from solution \cite{Kim85}. Along this surface normal the unit cell is composed of two molecular layers, each of which exposes an equivalent surface termination. Moreover, as the strong hydrogen bonds form in this orientation within a molecular layer, growth and dissolution is likely to proceed via single layers. This allows to consider step edges of the height of a single molecule \cite{Snyder09}, which considerably reduces the complexity of the model.

From the various possible step edge directions at this surface, we consider only the two close-packed edges perpendicular to the [100] and [010] direction, as well as the edge perpendicular to the [1$\bar{1}$0] direction, which exposes a more open step termination but preserves the hydrogen bonds between the aspirin dimers. The considered step structures are displayed in Fig. \ref{fig:Step_edges}. As depicted in Fig. \ref{fig:Step_edges} (a), step defect structures are created in the periodic boundary condition simulation supercell ({\em vide infra}) by rearranging half of the molecules from the top layer of a perfect surface towards the bottom surface, thus creating four step edges of the same type in the simulation cell in total.

\begin{figure*}
  \includegraphics[width=0.78\textwidth,clip]{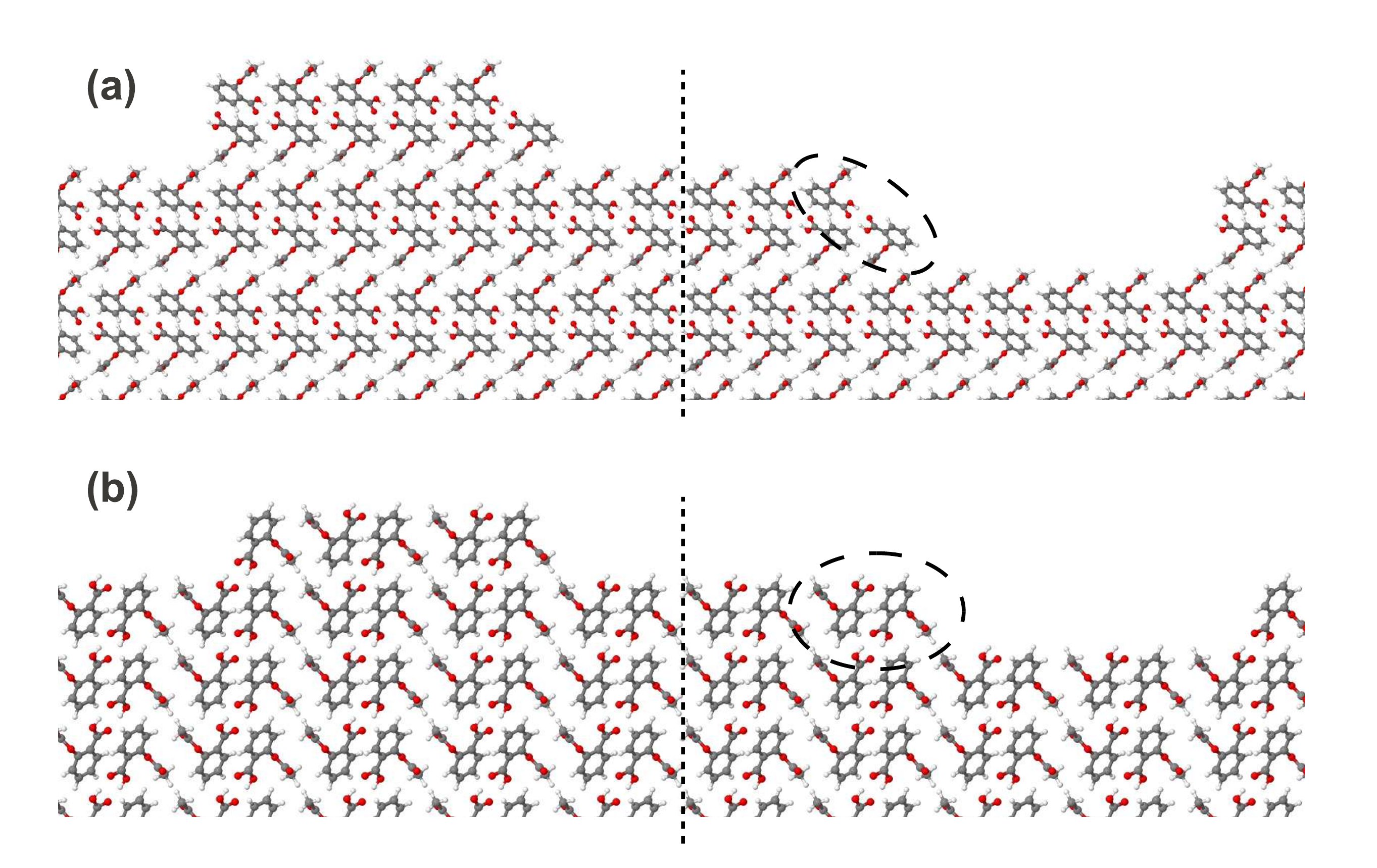}
  \caption{Stable kink structures at the (a) (100) and (b) (010) step edge. The separation into two partial systems is indicated by the dotted line (see text). The dimers that have been considered as dissolution units in the dissolution free energy calculations are marked by the dashed ellipses.}
  \label{fig:Kink_structures}
\end{figure*}

Kink sites are created along these three step directions. For each step direction, opposite edges, e.g. (100) respectively ($\bar{1}$00), expose slightly different terminations in terms of the terminal acetyl moieties pointing upwards into solution or downwards to the crystal surface. In order to ensure a precise and well-defined calculation of defect formation free energies, kink sites are therefore always created along the same edge. As the symmetry of the aspirin unit cell offers several possibilities to create kink arrangements at a given step edge, we define, with the ultimate scope of dissolution kinetics, the following criteria to reduce the number of structures in the free energy simulations: (1) The kink structure must be metastable on the time scale of the free energy calculations, i.e. the next possible molecular detachment process must be a rare event. (2) The detachment or attachment of a dissolution/growth unit must result in the same type of kink site to enable a continuous, self-sustaining process. (3) The edge termination between the newly created kink sites before and after re-arrangement must be the same, in order to avoid a systematic dependence of the defect formation free energy on the number of re-arranged molecules. While criteria (2) and (3) can be assessed by an inspection of the static crystal structure, the first point requires additional dynamic simulations. Specifically, we performed MD simulations over more than 20 ns to detect possible spontaneous detachment events and thereby assess the metastability of possible kink structure candidates. The kink structures that fulfill the criteria for the given step edges are shown in Fig. \ref{fig:Kink_structures}.

The transfer of molecules from the interface to bulk solution is investigated with the two stable kink structures at the (100) and (010) step edges as starting structures. In detail, we consider for both step directions the detachment of a pair of aspirin molecules from a centrosymmetric initial arrangement as dissolution unit, as depicted in Fig. \ref{fig:Kink_structures}. The dimer at the (100) edge is held together by a pair of hydrogen bonds, while the selected unit within the (010) step does not feature such intermolecular bonds among the two dissolving molecules. Due to their symmetry center both kink sites form half-crystal positions \cite{Kuvadia2011}, and the detachment of the selected molecules results in exactly the same type of kink site, thus allowing for a continuous dissolution or growth process via these processes. In both cases the final state, i.e. dissolved aspirin molecules, is the same.

In general, the notion of a perfect structure denotes that solely defect types of higher dimension than the type under
consideration are present, i.e. the reference arrangement of a kink site is a perfect step edge, while the reference structure of a step defect is a perfect crystal surface. As we are interested in the free energy change associated with single defect types only, we carefully create the surface structures by ensuring that initial and final model presents the same amount and structure of possible larger scale defects. We thus avoid any systematic dependence of the defect formation free energy on the width of the terrace in between two step edges, or the length of the edge in between two kink sites.

For all interface simulations we employ a slab model and a supercell geometry with periodic boundary conditions in all directions. The surface models are built from a unit cell, optimized using the generalized AMBER force field. The crystal slab thickness amounts to four unit cells, i.e. eight molecular layers. The interface systems are prepared by filling the vacuum gap between opposite slab surfaces with pre-equilibrated water molecules. The thickness of the water layer is chosen sufficiently large in order to retain a significant region of bulk-like water molecules in between the two slab surfaces, which is assessed by monitoring the water density profile. After an initial geometry optimization to remove spurious particle overlaps, and a short pre-equilibration run of 300\,ps MD simulations at constant volume and temperature, the correct height of the simulation cell is adjusted in a long 3\,ns simulation at constant pressure of 1\,atm in direction of the surface normal. In all simulations the lateral cell vectors remain fixed to the values dictated by the underlying crystal lattice. To simulate a single dissolved aspirin molecule in bulk water, we employ a cubic simulation cell with an initial volume of (5.0 $\times$ 5.0 $\times$ 5.0) nm$^3$. After 3\,ns of equilibration under isotropic pressure coupling to obtain the correct box size, the production simulations are carried out in the $NVT$ ensemble.

\subsection{Thermodynamic Cycle}

In this section we present the thermodynamic cycles, which we have devised in order to calculate the free energy difference between initial and final state. In the following, the initial state corresponds to the perfect crystal arrangement, while the final state presents the defect situation to be considered. As the free energy is a state function the cycle does not need to follow a physical pathway as long as all individual steps are carried out in a reversible way.
Our approach essentially follows the methodology that has been developed to simulate binding free energies in biological ligand/receptor complexes \cite{Woo,Wang2006}. The main part of these simulations comprises decoupling of the non-bonded interactions between the ligand and its surroundings (receptor, or solvent molecules, respectively) both in the initial and final state. The difference between the free energy changes of these two decoupling processes is then taken as the binding free energy. To ensure reversibility during the decoupling procedure, an additional set of restraints to the ligand molecule has to be introduced, in order to prevent it from abandoning its original position and orientation when the interactions are substantially reduced. The free energy cost of introducing as well as removing these restraints has to be taken into account in the total free energy balance.

In order to transfer the approach developed for ligand binding to surface defects at a molecular crystal/water interface, we identify the ligand with the set of surface molecules that have to be rearranged in order to create the desired defect structure. The thermodynamic cycle starts from the perfect crystal structure, equilibrated at the reference conditions of 300\,K temperature and 1\,atm pressure, by introducing the restraint potentials. As we are interested in preserving the positions as well as the orientation of the molecules to be shifted, we apply absolute position restraints via harmonic potentials on the atomic positions of all heavy atoms as well as the carboxyl hydrogen atoms of these marked molecules.
Additionally, we restrain each hydrogen atom belonging to the carboxylic acid moiety to preserve the structure of the hydrogen bond pairs. The reference positions for the involved atoms are taken as those of the perfect crystal molecules, with the center of mass of the entire set of atoms aligned to the equilibrium value obtained in a long $NVT$ MD simulation.
Switching on the position restraints gives rise to a free energy change of $\Delta G_{\rm perf}^{\rm restr}$. To avoid artifacts due to this translationally non-invariant restraint potential, we intentionally remove the translational symmetry by additionally tethering all atoms in either the bottom or central layer to their crystal positions throughout the entire free energy calculation, as described in section \ref{sec:simulation_details}.

With the restraint potentials switched on, the interactions between the restrained molecules and their surroundings. i.e. the remaining crystal molecules as well as the solvent, are decoupled. This process is accompanied by a free energy difference of $\Delta G_{\rm perf}^{\rm dec}$. In the completely decoupled state the restrained molecules can be considered to be virtually surrounded by vacuum. Hence, a shift of these molecules, while preserving their relative orientations as dictated by the position restraints, can be carried out at no additional free energy cost. Similarly, any symmetry operation on these molecules, such as e.g. rotation or point symmetry, contained in the space group of the crystal, can be performed at constant free energy. Based upon these considerations the set of molecules, including their restraint reference positions, is rearranged to create the desired final defect structure. In the new configuration the interactions of these molecules with the remaining part of the system are switched on again under position restraints,
yielding a free energy contribution of $-\Delta G_{\rm def}^{\rm dec}$. Finally, with the restrained molecules in the defect arrangement, the position restraints are removed to arrive at the final state. This step is accompanied by a free energy change of $-\Delta G_{\rm def}^{\rm restr}$. In cases where the resulting defect structure does not correspond to the reference pressure, the system volume has to be adjusted at a free energy cost of $\Delta G_{\rm vol}$.

The entire cycle therefore gives rise to a defect formation free energy change of
\begin{equation}
\Delta G = \Delta G_{\rm perf}^{\rm restr} + \Delta G_{\rm perf}^{\rm dec} - \Delta G_{\rm def}^{\rm dec} - \Delta G_{\rm def}^{\rm restr} + \Delta G_{\rm vol}\, .
\end{equation}

\subsection{Free Energy Calculations}

After outlining the general thermodynamic cycle to obtain the free energy differences between perfect and defect state,
we now describe the detailed methodology employed to calculate the individual free energy contributions.

The restraint free energy contribution in the step and kink free energy is obtained by thermodynamic integration of the negative generalized force $dG/dk = \langle dU/dk \rangle$ with respect to the spring constant of the position restraints $k$:
\begin{equation}
\Delta G^{\rm restr} = \int_0^{k_{max}} dk \left \langle \frac{ d U }{ d k } \right \rangle_{k} \, \quad .
\end{equation}
Here, $U$ denotes the potential energy of the system including restraint potentials. To this aim we perform a set of equilibrium simulations at various intermediate $k$-values between zero and the final restraint force constant,
calculating in each simulation the derivative of the restraint potential with respect to the force constant $k$, as
\begin{equation}
\left \langle \frac{ d H }{d k} \right \rangle_{k} = \frac{1}{2} \left \langle \left (\mathbf{R}_{\rm restr} - \mathbf{R}_{\rm restr}^0 \right )^2 \right \rangle_{k} \, ,
\end{equation}
where $\mathbf{R}_{\rm restr}$ denotes the vector comprising the position of all restrained atoms, and the superscript 0 refers to the respective reference positions. In practice, for the simulations presented in this study, thermodynamic integration is carried out numerically by using the trapezoidal rule, based on seven $k$-values of 0, 50, 100, 250, 500, 750, and 1000 kJ/mol/nm$^2$. To sample the steep gradient at small $k$ values, a finer spacing is employed in this region.
According to initial test simulations a final value of $k_{\rm max}$ = 1000.0 kJ/mol/nm$^2$ has been found to provide a satisfying balance between preserving the molecular arrangement and a reasonably low restraint free energy, associated with accordingly low error values. The total error of the restraint free energy is calculated via error propagation
of the statistical uncertainties of the individual generalized force values.

For the simulation of the decoupling free energy, a switching parameter $\lambda = (\lambda_{\rm el},\lambda_{\rm LJ})$ is employed to interpolate between the original Hamiltonian and the auxiliary system without non-bonded interaction between the marked molecules and the remaining system. The two components of the $\lambda$-vector couple to electrostatic and Lennard-Jones interactions, respectively. According to the established protocols \cite{Wang2006}, we first switch off the electrostatic interactions by applying a linear coupling parameter $\lambda_{\rm el}$ to the electrostatic contribution of the potential energy:
\begin{equation}
U^{\rm el}(r_{ij},\lambda_{\rm el}) = (1-\lambda_{\rm el}) \frac{q_i q_j}{ r_{ij}}\, \quad .
\end{equation}
$q_{i,j}$ denotes the partial charge of atoms $j$ and $i$, the former being part of the decoupling set, while the latter belongs to the remaining part of the system. We generally employ intermediate $\lambda_{\rm el}$ values at a spacing of 0.1, as well as an additional point at $\lambda_{\rm el} = 0.05$, to improve sampling of the steep initial free energy gradient.

Subsequently, the Lennard-Jones interactions are switched off with the electrostatic interactions remaining in the $\lambda_{\rm el} = 1.0$ state. To avoid convergence issues due to possible particle overlaps at partially switched off repulsive interactions, the diverging Lennard-Jones potentials are continuously transformed into soft-core potentials, as
\begin{equation}
U^{\rm LJ}(r_{ij},\lambda_{\rm LJ}) = (1 - \lambda_{\rm LJ}) U_{ij}^{\rm LJ} ( \alpha \sigma_{ij}^6 \lambda_{\rm LJ} + r) \, \quad .
\end{equation}
Here, $\sigma_{ij}$ denotes the standard Lennard-Jones parameter, $\alpha$ is a soft-core parameter that has been set to the GROMACS default value of 0.3, and $U_{ij}^{\rm LJ}$ represents the original Lennard-Jones potential between particles $j$ and $i$. We employ basic intermediate $\lambda_{\rm LJ}$ values at a spacing of 0.1. As the accuracy at large $\lambda_{\rm LJ}$-values is affected by the onset of penetration of solvent into the decoupled region, we place four to six additional $\lambda_{\rm LJ}$ points between 0.6 and 1.0, depending on the size of the defect structure. Each state is simulated for at least 2\,ns, after an initial equilibration period of 1\,ns. The free energy differences between neighbouring $\lambda$-states are calculated using the Bennet-acceptance-ratio (BAR) method \cite{Bennet76}, as implemented in the GROMACS package. The same procedure is repeated for the restrained defect arrangement.

For the transfer of individual molecules from the crystal surface to bulk solution a slightly different scheme is employed.
Due to the substantially different translational and rotational entropy in initial and final state, position restraints on these degrees of freedom are imposed separately, as described in the following: At first the center-of-mass (COM) vector of the marked molecule is restrained via a harmonic potential with a spring constant of $k^{\rm com} = 10000.0$ kJ/mol/nm$^2$ to its average position, as obtained in an equilibrium simulation. Subsequently, the mean-square-displacement (MSD) between the atomic positions relative to the molecular COM, and those of the perfect crystal molecule, 
\begin{equation}
\xi_{\rm rel}^2 = [(\mathbf{R_{\rm restr}}-\mathbf{R_{\rm com}}) - (\mathbf{R_{\rm restr}^0}-\mathbf{R_{\rm com}^0})]^2 \, , 
\end{equation}
is restrained by a quartic potential $U^{\rm msd} = 0.5 \, k^{\rm msd} \, \xi^4$. Again, we consider all heavy atoms as well as the carboxy hydrogen. Technical reasons, in detail the implementation of the MSD within the PLUMED code, dictate the use of the quartic form. Based on initial test simulations, a value of $k^{\rm msd}=100000$ kJ/mol/nm$^4$ has been determined as suitable to maintain the molecular equilibrium MSD fluctuation even in the entirely decoupled system.

The restraint free energy of the COM and the MSD of single molecules is calculated via free energy perturbation, as
\begin{equation}\label{eq:ThermPert}
\exp \left ( -\beta \Delta G^{\rm COM,MSD} \right ) = \langle \exp \left (-\beta U^{\rm COM,MSD} \right ) \rangle_{k^{\rm restr}=0} \, \quad .
\end{equation}
Prior to transferring the decoupled molecule into water solution, the restraint on the MSD is switched off in vacuum,
as the absence of solvent molecules allowed to access longer simulations times and thus an enhanced sampling of molecular orientations to obtain a converged $\Delta G_{\rm vac}^{\rm msd}$ value. The solvation free energy is then obtained by decoupling the interactions of the aspirin molecule with the solvent molecules in the same way as described above.
Finally, the release of the COM-restraint into bulk solution at standard state concentration $c_0$ can be evaluated analytically, as \cite{Wang2006}
\begin{equation}
e^{-\beta \Delta G_{\rm sol}^{\rm com}} = c_0 \int_{V_0} d \mathbf{r}_{\rm com} e^{-\beta \frac{k}{2} \mathbf{r}_{\rm com}^2} \, \simeq \, c_0 \left ( \frac{2 \pi k_B T}{k} \right )^2 \, \quad .
\end{equation}
This relation implies that the solution concentration of the dissolved molecules changes the corresponding chemical potential only via contributions of translational entropy. We thus exploit the common approximation of the activity by the concentration. For the case of aspirin we justify this approximation by the comparably low solubility concentration $c_{\rm sat} \simeq 0.02$\,mol/l \cite{Perlovich,Wen2004,Edwards}.

We note that all free energy calculations in this work are carried out at constant volume. In all cases where the defect structure is obtained from the perfect structure by a mere rearrangement of molecules within the same simulation cell, both systems typically occupy almost exactly the same equilibrium volume at reference pressure, thus avoiding the need for a volume free energy correction. If the initial and final state are simulated using different systems, the process of decoupling the interactions of a set of molecules at constant volume results in a pressure change $\Delta p$. In order to recover the reference state, the reference pressure $p_0$ has to be restored by changing the volume of the system, which is accompanied by a free energy change $\Delta G_{\rm vol}$. As an exact pressure value is difficult to obtain from the simulations in the presence of position restraints, we estimate the pressure difference upon a volume change of $\Delta V$ via $\Delta p = - \Delta V /(V \kappa)$. Here $\kappa$ denotes the compressibility of the system, which is assumed constant for small $\Delta V /V$. Accordingly, the free energy change to recover the reference pressure $p_0$ after decoupling the marked molecules can be estimated as
\begin{equation}\label{eq:DeltaG_vol}
\Delta G_{\rm vol} = \int_{V_1}^{V_0} dV (p_0 + \Delta p(V)) \simeq p_0 V_0 - \frac{1}{2} \frac{ \Delta V^2 }{V_0 \kappa} \, \quad ,
\end{equation}
$V_1$ being the fixed system volume during the decoupling simulations and $V_0$ the volume equilibrated at reference conditions after decoupling. The detailed evaluation of Eq. \ref{eq:DeltaG_vol}, if necessary, shall be discussed in the context of the results of the respective defect formation free energy simulations.

\section{Results}

\subsection{Step Edge Defect}

\begin{table*}[t]
\caption{\label{tab:FE_Step}
Free energy components during the calculation of step free energies for different edge orientations at the aspirin(001)/water interface, as well as final step free energies per step length.}
\begin{tabular}{|l|l||c|c|c|}
\hline
 System & Component & (100) & (010) & (1$\bar{1}$0) \\
\hline
Perfect & $\Delta G_{\rm perf}^{\rm restr}$ [kJ/mol] & 472.1 $\pm$ 1.2 & 481.1 $\pm$ 1.2 &  1887.9 $\pm$ 4.9  \\
        & $\Delta G_{\rm perf}^{\rm dec}$ [kJ/mol]  & 9037.7 $\pm$ 3.8 &  9046.5 $\pm$ 3.2 & 8233.7 $\pm$ 6.1  \\
Step    & $\Delta G_{\rm def}^{\rm restr}$ [kJ/mol] & 627.3 $\pm$ 1.7 &  630.9 $\pm$ 1.6 &  2013.5 $\pm$ 5.1  \\
        & $\Delta G_{\rm def}^{\rm dec}$ [kJ/mol] & 8510.3 $\pm$ 3.1 & 8509.7 $\pm$ 4.0 & 7381.7 $\pm$ 3.9  \\
\hline
\multicolumn{2}{|c|}{$\Delta G^{\rm step}$ [kJ/mol/nm]} & 12.07 $\pm$ 0.2 & 14.11 $\pm$ 0.13 &  17.32 $\pm$ 0.24 \\
\hline
\end{tabular}
\end{table*}

\begin{figure*}
  \includegraphics[width=0.6\textwidth,clip]{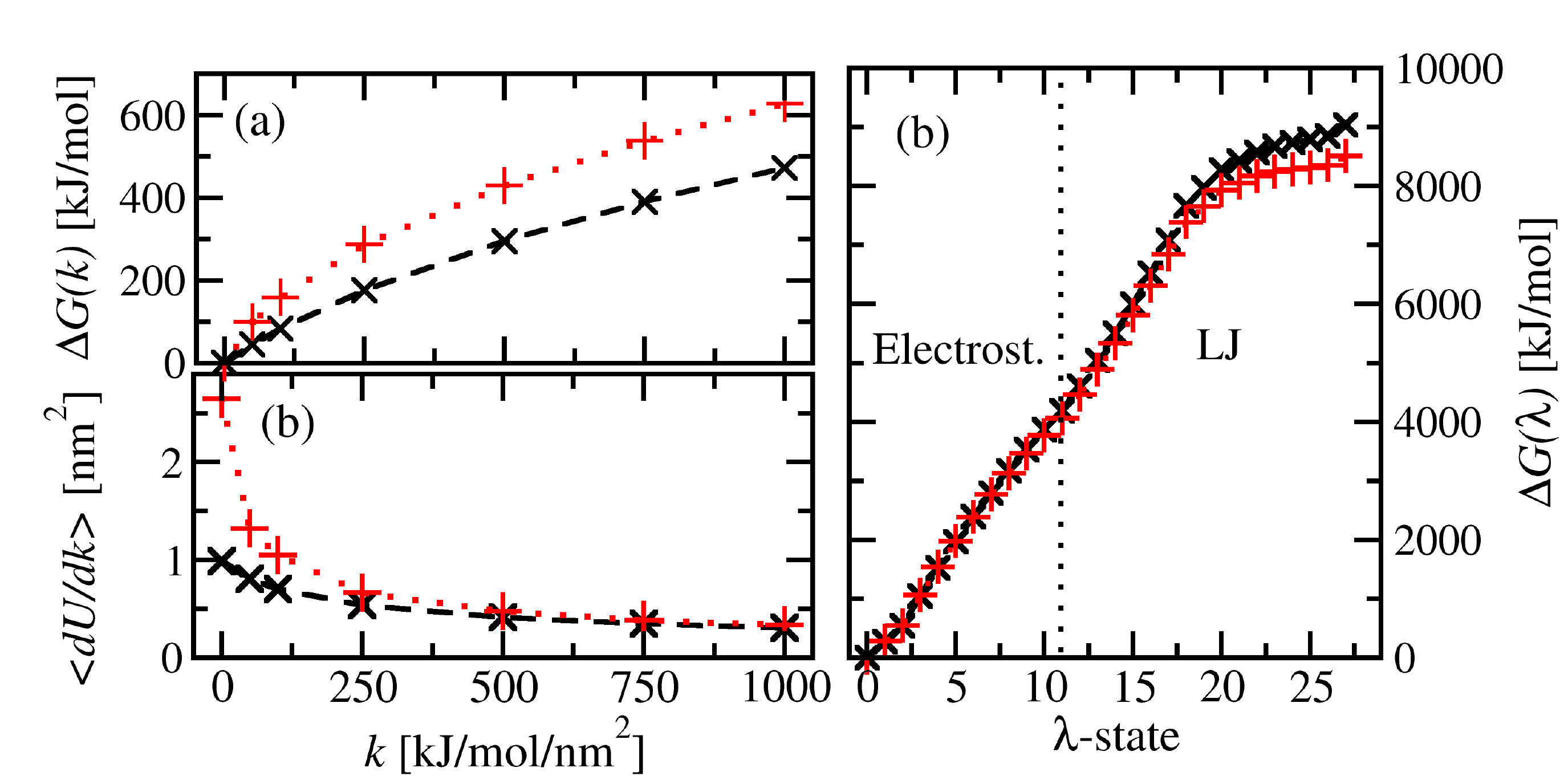}
  \caption{Free energy contributions in the defect formation free energy of the (100) step edge:
(a) Integrated restraint free energy as a function of the spring constant $k$. 
(b) Negative generalized force $\langle dU/dk \rangle$ on the spring constant $k$.
(c) Free energy change upon decoupling the interactions of the marked molecules with the remaining system.
The symbols refer to the perfect surface (x, dashed lines) and the step structure (+, dotted line), respectively.}
  \label{fig:FE_step_100}
\end{figure*}

We start our investigation by considering the free energy cost of creating a step defect along a chosen direction at the aspirin (001)/water interface. The free energy contributions of the steps in the thermodynamic cycle are listed in table \ref{tab:FE_Step}. Considering the case of the (100) step edge, the details of the restraint free energy calculation are displayed in Fig. \ref{fig:FE_step_100} (a,b), in terms of the generalized force and the integrated free energy along the spring constant $k$. The generalized force decreases monotonically with increasing $k$-value. The initial steep slope is sufficiently captured by the finer spacing of intermediate points in this region, while fewer points are necessary to interpolate the following part of the curve at larger spring constant values. We note that the curves associated with the perfect and the defect surface show qualitatively the same behavior, although the restraint free energy is larger at the step surface, as the molecules within the step edge possess enhanced rotational and vibrational freedom compared to molecules in the pefect surface. 

The accumulated free energy differences during the decoupling process using a total number of 28 $\lambda$-states are displayed in Fig. \ref{fig:FE_step_100} (c). The first section ($\lambda$ states 0 - 11) reveals a monotonic increase in free energy, as this region represents the decoupling of the electrostatic interactions. The remaining part describes the decoupling of the Lennard-Jones interactions. As particularly the last section is affected by pronounced structural re-ordering processes, particularly via water molecules entering the location of the marked defect molecules, the curve does not reveal a clear monotonic behavior any longer. While the decoupling free energies of perfect and defect system reveal very similar behavior when switching off the electrostatic as well as the initial part of the Lennard-Jones interactions, more pronounced free energy differences arise in the final part of the Lennard-Jones region, in particular in the transition region between steep increase and plateau-like behavior.

We note that the absolute values of the free energy contributions are large. This must be attributed to the fact that a relatively large number of molecules is involved in the decoupling as well as restraining process. Yet, the major part of these free energies cancels out in the free energy differences. The absolute statistical error values remain reasonably low, even compared to the free energy differences, due to the careful choice of intermediate states. In total, we arrive at step free energy values per edge length of $\Delta G^{\rm step}$ = 12.07 kJ/mol/nm for the (100) orientation, 14.11 kJ/mol/nm for the (010) orientation, and 17.32 kJ/mol/nm for the (1$\bar{1}$0) orientation. Exposing a close-packed edge termination while preserving all hydrogen bond pairs, the (100) step edge thus provides the lowest free energy difference, and must therefore be considered as thermodynamically most stable step edge. The other close-packed edge, i.e. at the (010) step, possesses a similarly small step free energy per length. The open (1$\bar{1}$0) termination, however, reveals a considerably larger value, thus being the least stable among the considered step edges.

For comparison we have calculated the step energy based on the rigid crystal approximation for the two most stable edge directions. Considering a single layer of molecules, we calculate the potential energy difference between perfect layer and two half-layers cleaved along the respective step edge. The vacuum step energy is afterwards corrected by the difference in solvation energies between perfect and defect arrangement. Using this methodology, we arrive at a vacuum step energy of 36.9 kJ/mol/nm for the (100) step and 58.8 kJ/mol/nm for the (010) step. The solvation correction amounts to -6.9 kJ/mol/nm, respectively -20.9 kJ/mol/nm, which gives rise to total step energies of 30.0 kJ/mol/nm for the (100) step, and 37.9 kJ/mol/nm for the (010) step edge. These values fall in the same order of magnitude as the MD results, but exceed them by a factor between two and three. The ratio between both step directions is approximately the same for both approaches, which may be the reason why the rigid crystal approximation still yields in many cases \textit{relative} crystal shapes, which are in good agreement with experiments \cite{Lovette12, Kuvadia2011}.

\subsection{Kink Free Energy}

After the step defects we turn towards the consideration of kink sites within a given step edge. The chosen kink rearrangement, as displayed in Fig. \ref{fig:Kink_structures} requires a considerable length of the edge within the simulation cell to avoid finite size effects due to interactions between neighboring kink sites. Accordingly, the resulting system size poses large computational demands. However, as during all free energy simulations half of the system is idle, we have chosen to split the entire system into two separate simulation cells, each containing half of the step edge, as depicted in Fig. \ref{fig:Kink_structures}. The transfer of the decoupled set of molecules from the perfect edge towards forming a row of ad-molecules takes places across different systems in this case, which can still be carried out at constant free energy. Nevertheless, one has to take care that the combined initial and final system represents the correct thermodynamic state, in particular the same reference pressure $p_0$. To estimate the free energy upon restoring the reference pressure by decreasing the volume of the initial system after decoupling, we considered the necessary volume change $\Delta V_{i}$. Similarly, before switching on the interactions of the shifted row of ad-molecules in the final system, the volume is increased by $\Delta V_{f}$ in order to recover the reference pressure after introducing these molecules. Both volume changes have been determined accurately in separate simulations at constant pressure, resulting in almost exactly the magnitude $\Delta V_{i} \approx -\Delta V_{f}$. Employing Eq. \ref{eq:DeltaG_vol} and assuming that both systems have the same compressibility value due to their almost identical compositions, both free energy contributions can be assumed to cancel to the largest extent. Accordingly, we have not included any volume contributions into the total kink free energy balance. In fact, a test calculation of a large step edge to simulate both parts simultaneously, thus avoiding the need for a volume correction, has revealed the same kink free energy within the statistical uncertainties, compared to the separate treatment of the smaller systems.

\begin{table*}[t]
\caption{\label{tab:FE_Kink}
Free energy components during the calculation of kink free energies for kinks within step edges of several orientations, as well as final kink free energies per kink site.}
\begin{tabular}{|c||c|c|c|c||c|}
\hline
System & $\Delta G_{\rm perf}^{\rm restr}$ [kJ/mol]  & $\Delta G_{\rm perf}^{\rm dec}$ [kJ/mol]  & $\Delta G_{\rm def}^{\rm restr}$ [kJ/mol]  & $\Delta G_{\rm def}^{\rm dec}$ [kJ/mol] & $\Delta G^{\rm kink}$ [kJ/mol]  \\
\hline
(100)         & 93.6 $\pm$ 0.5 & 1308.4 $\pm$ 1.4 & 102.9 $\pm$ 0.6 & 1244.5 $\pm$ 1.4 & 13.7 $\pm $ 0.6 \\
($\bar{1}$00) & 106.0 $\pm$ 0.6 &  1267.1 $\pm$  1.3 & 115.3 $\pm$ 0.7  & 1198.3 $\pm$  1.6 & 14.9 $\pm$ 0.6 \\
(010)         & 55.6 $\pm$ 0.4 & 574.8 $\pm$ 1.0 & 56.2 $\pm$ 0.4 & 557.3 $\pm$ 1.2 & 4.2 $\pm$ 0.5 \\
(0$\bar{1}$0) &  56.1 $\pm$ 0.4 & 572.3 $\pm$ 2.7 & 61.8 $\pm$ 0.5 &  553.9 $\pm$ 2.2  & 3.2 $\pm$ 0.9 \\
(1$\bar{1}$0) & 61.6 $\pm$ 0.4 & 812.1 $\pm$ 2.4 & 67.4 $\pm$ 0.5 & 800.5 $\pm$ 0.8 & 1.5 $\pm$ 0.7 \\
($\bar{1}$10) & 75.0 $\pm$ 0.5 & 798.9 $\pm$ 3.0 & 85.8 $\pm$ 0.6 &  784.6 $\pm$ 1.8  & 0.9 $\pm$ 0.9 \\
\hline
\end{tabular}
\end{table*}

The contributions of the kink free energies for kink sites at the two different terminations of the (100), (010), and (110) edges are summarized in table \ref{tab:FE_Kink}. In general, the details of the restraint respectively decoupling free energy calculations follow qualitatively the behavior encountered for the step free energy calculations and are therefore not further discussed. For each step direction the kink free energy yields values of similar order of magnitude for both edge terminations. A comparison between different edge directions, however, reveals more pronounced differences. The most stable edge orientations, i.e. the (100) and ($\bar{1}$00) steps, yield comparably large kink free energies in the range 5 to 7 $k_B T$, whereas the least stable (1$\bar{1}$0) and ($\bar{1}$10) edges provide very low free energy values, much smaller than the thermal energy. Kinks along the (010) and (0$\bar{1}$0) step edges yield defect formation free energies in between these two extreme cases with a magnitude of about 1 - 2 $k_B T$.

\subsection{Dissolution Free Energy}

\begin{table*}[t]
\caption{\label{tab:FE_Dissolution}
Free energy components during the calculation of dissolution free energies from different defect sites at the aspirin(001)/water interface. 
The vacuum and solution contributions are the same for all defect types and thus only reported for the first case. 
The final $\Delta G^{\rm diss}$ values always refer to the free energy change per aspirin molecule.}
\begin{tabular}{|l|l||c|c|c|}
\hline
 System & Component & 2-mol. Kink (100) & 2-mol. Kink (010) \\
\hline
Interface & $\Delta G_{\rm inter}^{\rm com}$ [kJ/mol] & 8.855 $\pm$ 0.001 & 10.55  $\pm$ 0.45 \\
        & $\Delta G_{\rm inter}^{\rm msd}$ [kJ/mol]   & 0.21 $\pm$ 0.05   & 0.304  $\pm$ 0.002 \\
        & $\Delta G_{\rm inter}^{dec,1}$ [kJ/mol]     & 108.16 $\pm$ 0.3 & 112.4 $\pm$ 0.6 \\
        & $\Delta G_{\rm inter}^{dec,2}$ [kJ/mol]     &  99.15 $\pm$ 0.4 &  94.4 $\pm$ 0.3 \\
Vacuum   & $\Delta G_{\rm vac}^{\rm msd}$ [kJ/mol] & 16.25 $\pm$ 0.5 &  \\
 Solution & $\Delta G_{\rm sol}^{\rm dec}$ [kJ/mol] & 57.2 $\pm$ 0.3 &  \\
         & $\Delta G_{\rm sol}^{\rm com}$ [kJ/mol] &  25.43 &   \\
\hline
\multicolumn{2}{|c|}{$\Delta G^{\rm diss}$ [kJ/mol]} & 9.31 $\pm$ 0.5 & 9.65 $\pm$ 0.7 \\
\hline
\end{tabular}
\end{table*}

Finally, the free energy change of transferring molecular dimers from the stable kink sites at the crystal/water interface into bulk solution is calculated. The COM-, respectively MSD-restraint potentials are imposed simultaneo usly to both kink site molecules. With the restraint potentials fully applied, the decoupling process is performed in a successive manner, starting with the more exposed molecule. The free energy contributions of the different intermediate steps for the two considered kink sites are listed in table \ref{tab:FE_Dissolution}. We assess the validity of the free energy perturbation technique to calculate $\Delta G_{\rm inter}^{\rm com}$ and $\Delta G_{\rm inter}^{\rm msd}$ by comparing the exponential average in Eq. \ref{eq:ThermPert} taken at $k=0.0$ and at the final value $k=k_{\rm com}$, respectively $k=k_{\rm msd}$ from 5 ns of simulation time in each state, which yields consistent results. The free energy difference upon releasing the MSD restraint potential, i.e. $\Delta G_{\rm vac}^{\rm msd}$, is calculated by averaging over a long vacuum simulation trajectory of 80\,ns. Here, the quality of conformational sampling is assessed by comparing the free energy profile, calculated from the histogram of the accessed MSD values via $G(\xi^2) = -k_B T \ln(P(\xi^2))$,
to the corresponding converged free energy profile, obtained in 40\,ns of well-tempered metadynamics simulations with $\xi^2$ as collective variable. Both curves reveal very good agreement, and the restraint free energy calculated by integrating along the exponential free energy, as described in Refl. \onlinecite{Wang2006}, agrees well within the statistical uncertainties with $\Delta G_{\rm vac}^{\rm msd}$ as calculated from the equilibrium simulation. The pronounced difference of this value compared to $\Delta G_{\rm inter}^{\rm msd}$ reflects the substantially increased rotational entropy after release into an isotropic environment.Similarly, releasing the COM-restraint into a standard state concentration of 1 mol/l, gives rise to a translational entropy contribution of $- \Delta G_{\rm sol}^{\rm com} = -25.43$ kJ/mol, thus exceeding the corresponding interface values $\Delta G_{\rm inter}^{\rm com} = 4.4$, respectively $5.3$ kJ/mol per molecule by far. As the total free energy differences associated with the interface systems correspond to two aspirin molecules, the value is divided by 2 to obtain the molecular free energy values. Subsequently, the remaining vacuum and solution contributions are subtracted, to yield final dissolution free energies per molecule of 9.31 and 9.65 kJ/mol for the (100) and (010) step edge, respectively. Both values agree within the uncertainties, as expected for centrosymmetric dissolution units at half-crystal positions \cite{Kuvadia2011}.

Since decoupling the interactions with the transferred aspirin molecules at the interface as well as in bulk solution at constant volume results in a decrease in pressure, the volumes of both systems have to be adjusted accordingly to restore the standard pressure. In the following we evaluate the corresponding free energy correction for the case of a kink at the (100) step edge, while noting that the behaviour at the (010) step is essentially the same. The total volume change after $NPT$ equilibration in the final interface system amounts to $\Delta V_{\rm inter} = -0.47$ nm$^3$, while the corresponding value in bulk solution calculates as $\Delta V_{\rm sol} = -0.26$ nm$^3$. We note that the former value refers to the transfer of two molecules, while the latter value includes only a single molecule. Applying Eq. \ref{eq:DeltaG_vol} with the experimental compressibility of water $\kappa_{\rm water}=4.5 \cdot 10^{-5}$ bar$^{-1}$ and taking as compressibility of the interface system the value of $\kappa_{\rm inter} = 3.8 \cdot 10^{-5}$ bar$^{-1}$, calculated from the simulations as described in the appendix, we arrive at values of $\Delta G_{\rm inter}^{\rm vol} = -0.91$ kJ/mol, and $\Delta G_{\rm sol}^{\rm vol} = -0.3$ kJ/mol. By normalizing the interface value to one molecule in order to obtain the molecular free energy difference, the correction term amounts to \mbox{$\Delta G_{\rm inter}^{\rm vol}/2 - \Delta G_{\rm sol}^{\rm vol} = 0.15$} kJ/mol. This value is considerably smaller than the total statistical error of the dissolution free energy, thus rendering the volume correction negligible.

\section{Thermodynamic Considerations and Discussion}

\begin{figure}
  \includegraphics[width=0.49\textwidth,clip]{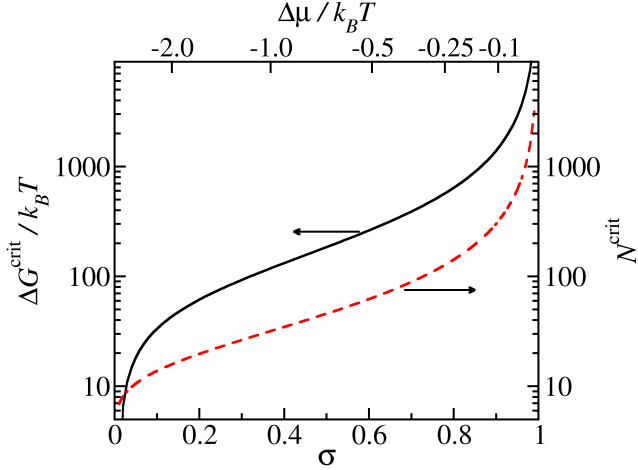}
  \caption{Thermodynamic free energy barrier $\Delta G^{\rm crit}$ and number of molecules $N^{\rm crit}$ associated with the critical pit nucleus, as obtained by a Wulff construction at undersaturation $\sigma = c_{\rm sol}/c_{\rm sat}$.}
  \label{fig:2DNucleation}
\end{figure}

Having established the reliable calculation of defect formation free energies, we shall now proceed towards a thermodynamic interpretation of the acquired free energy values with respect to their application to predict crystal growth and dissolution processes. For clarity we will discuss the results primarily in the context of crystal dissolution, although all the concepts may be transferred in an analogous way to growth. We begin by considering 2D-nucleation, via the formation of pits (dissolution) or islands (growth). According to classical nucleation theory, the formation free energy of such a 2D-structure from a perfect surface calculates as the balance of the chemical potential difference $\Delta \mu \simeq k_B T \ln(\sigma)$ between a molecule in the crystal and in bulk solution at a given relative saturation $\sigma = c_{\rm sol}/c_{\rm sat}$, and the free energy cost for creating the surrounding edges of length $l_i$ \cite{Lovette12}:
\begin{equation}\label{eq:DeltaG_nucl}
\Delta G^{\rm pit} = \Delta n_{\rm cryst} \Delta \mu +  \sum_{i=1}^{N_{\rm steps}} l_i \Delta G_{i}^{\rm step} \, \quad.
\end{equation}
Here, $\Delta n_{\rm cryst}$ denotes the change in the number of crystal molecules upon creating the 2D defect and $N_{\rm steps}$ is the number of step edges taken into account. The optimum shape of an island or pit can be determined by employing a 2D-Wulff-construction \cite{Wulff}, as described in Ref. \onlinecite{Lovette12} to minimize the total edge free energy. Inserting the calculated step free energies for the (100), (010), and (1$\bar{1}$0) edges, we obtain an optimum edge length ratio of 1:0.82:0.08, meaning that pits are most likely to be bounded by the two close-packed step edges. The (1$\bar{1}$0) and ($\bar{1}$10) edges, on the contrary, contribute only little to the shape. From the edge lengths the critical size of the pit as well as the associated thermodynamic free energy barrier can be calculated \cite{Lovette12}. The resulting values $\Delta G^{\rm crit}$ and $N^{\rm crit}$ are displayed in Fig. \ref{fig:2DNucleation} as a function of the chemical potential difference $\Delta \mu$, respectively the corresponding undersaturation $\sigma$. For most undersaturation values the free energy barrier is found to be extremely large, exceeding 100 $k_B T$, and thus rendering the observation of 2D nucleation thermodynamically unlikely. Only at very pronounced undersaturation conditions $\sigma < 0.05$ or $\Delta \mu < -3.0 k_B T$, the nucleation free energy barrier is reduced to values below 10 $k_B T$. In this regime the predicted critical nucleus comprises less than ten molecules, which may limit the validity of the employed continuum approach.

These results suggest that under moderate undersaturation conditions spiral-like dissolution patterns may rather be dominant, emanating from screw dislocations at the interface as proposed in Ref. \onlinecite{Burton51,Cabrera}. In this process the dissolution rate is determined by the displacement velocities of the step edges exposed during the rotation of the spiral. Based on the corresponding theoretical framework \cite{Snyder09,Lovette12}, the edge velocity can be estimated by the net flux of molecules into kink sites and the density of kink sites along the respective step edge. While the former quantity can not be entirely reduced to equilibrium properties of the system, but requires accelerated MD simulations as presented in Ref. \onlinecite{Schneider14b}, the kink site density can be estimated from the corresponding kink free energies. Assuming quasi-equilibrium conditions in proximity of the step edge, the probability of encountering at a given molecular site either an outward or an inward kink rather than a straight continuation can be calculated from the kink free energy as \cite{Burton51}
\begin{equation}\label{eq:kink_dens}
\rho_{\rm kink} = \bar{n}^{-1} = \left [ 1 + \frac{1}{2} e^{\Delta G^{\rm kink}/k_B T} \right ]^{-1} \, \quad ,
\end{equation}
where $\bar{n}$ denotes the average number of molecules between two kink sites. Inserting the free energy values acquired in our extensive MD simulations, we arrive at average kink separations of 122 and 197 molecules for the (100) and ($\bar{1}$00) step edge, 3.7 and 2.8 molecules for the (010) and (0$\bar{1}$0) edges, and 1.9 and 1.7 for the (1$\bar{1}$0) and ($\bar{1}$10) step directions. These findings indicate that the (100) and ($\bar{1}$00) step directions must be expected to expose a comparably smooth edge termination with only few kink sites, at which growth and dissolution can take place.
On the contrary the (1$\bar{1}$0) and ($\bar{1}$10) step directions possess kink free energies smaller than $k_B T$ and should therefore exhibit an extremely high equilibrium density of kink sites. The corresponding edges will thus rather reveal a rough appearance than a well-defined edge termination. Dissolution and growth at this edge is correspondingly not restricted to few sites along the edge, but can take place essentially at all molecular sites along the step edge independent of their particular defect type. The kink-limited growth and dissolution framework is therefore not applicable to the open step directions. The (010) respectively (0$\bar{1}$0) step directions reveal defect densities which can not be clearly attributed to one or the other regime.

The standard state dissolution free energy of single kink site units characterizes the equilibrium between these kink sites and the dissolved state in bulk solution. The dissolution of the centrosymmetric dimer units can be related to the solubility equilibrium. Assuming that dissolution takes place primarily at kink sites, the net flux of molecules out of, respectively into these kink sites, as described by the equation
\begin{equation}
{\rm ASA}_2^{\rm kink} \leftrightarrow 2 {\rm ASA}^{\rm sol} \, \quad .
\end{equation}
vanishes at equal chemical potential, i.e. at saturation conditions, and then results in no overall growth or dissolution of the crystal. The corresponding solubility equilibrium can be written as \cite{Stack12}
\begin{equation}
K_{\rm eq} = e^{-2 \Delta G_0^{\rm diss}/k_B T} = \frac{ a_{\rm sat}^2 }{ a_{\rm kink} } \, \quad .
\end{equation}
As the detachment of centrosymmetric units from a kink site reproduces exactly the same defect site, the corresponding kink activity is set to 1 for a batch dissolution reaction. For the reasons discussed above, the activity of the dissolved molecules is approximated by their solution concentration relative to the chosen standard state concentration of 1 mol/l. Under these assumptions the solubility concentration of aspirin in water environment can be estimated as $c_{\rm sat} = c_0~e^{-\Delta G_0^{\rm diss}/k_B T} = 0.024 \pm 0.005$ and $0.021 \pm 0.006$ kJ/mol considering the (100) and the (010) kink site respectively. These values are in very good agreement with experimentally measured solubility concentrations in water at 298 K, ranging from 0.0178 \cite{Perlovich} via 0.021 \cite{Wen2004}, to around 0.025 mol/l \cite{Edwards}. This agreement demonstrates that our free energy scheme is suitable for an {\em in-silico} prediction of crystal solubilities and highlights the good performance of the GAFF to reproduce the energetics of the aspirin/water interface. It moreover indicates an excellent accuracy of the proposed method for defect formation free energy calculations at the crystal/water interface in general.

\section{Conclusions}

We have presented a free energy simulation scheme to accurately calculate defect formation free energies at the aspirin/water interface. In contrast to prevalent rigid-crystal methods to estimate defect formation free energies, the proposed technique allows to take into account explicit solvation, finite temperature, as well as anharmonic effects. By employing a thermodynamic cycle comprising a combination of restraining and decoupling, together with a suitable stratification scheme we obtain reproducible results with an excellent control of the statistical errors. As a showcase, we have considered step and kink defects at the aspirin (001)/water interface. Based on the calculated step free energies we expect that the most stable edges are the close packed (100) and (010) directions, whereas the (1$\bar{1}$0) direction plays a minor role at this surface. Applied to classical nucleation theory, the results furthermore suggest that the nucleation of pits during dissolution at moderate undersaturation conditions is accompanied by a huge thermodynamic free energy barrier, which renders the occurrence of such events unlikely. The acquired kink free energy values indicate that the (100) step, as well as its opposite edge, expose a comparably straight edge with only few kink defects, whereas the other considered edges exhibit a rather rough appearance with a higher density of defects.

When comparing the step free energy results to the values obtained within the rigid crystal approximation, we find that the latter approach produces a pronounced overestimation by a factor of two to three, even in the presence of an implicit solvent correction. Due to the neglect of entropic and finite temperature contributions, as well as an inaccurate description of solvent effects in the rigid crystal approximation, we expect a similar overestimation also for other defect types. This may not lead to dramatic errors in the prediction of \textit{relative} shapes and features, but it must be expected to have severe implications for the calculation of absolute growth or dissolution rates, as central quantities, such as kink site densities, depend exponentially on the respective defect formation free energies. The rigid approximation may therefore not provide sufficiently accurate values for a quantitative prediction of absolute dissolution or growth rates.

Finally, we have calculated the free energy difference associated with the dissolution of molecular dimers from two different kink sites. Due to their centrosymmetric properties both kink site units yield consistent values.
Moreover, the solubility concentration predicted based on the free energy difference agrees well with the range of experimental values from literature, which underlines the accuracy of the proposed free energy simulation scheme.
Combined with suitable simulation techniques for the accelerated calculation of detachment rate constants \cite{Schneider14b}, these defect formation free energies and the derived interfacial defect distributions may in a next step be used to formulate an {\em in silico} microkinetic model towards the quantitative prediction of absolute dissolution rate constants of pharmaceutically relevant organic crystals.

\section*{Acknowledgements}

We gratefully acknowledge support by the Deutsche Forschungsgemeinschaft (DFG) through grant RE1509/18-1.

\appendix

\section{Estimate of compressibility values}

\begin{figure}
  \includegraphics[width=0.5\textwidth,clip]{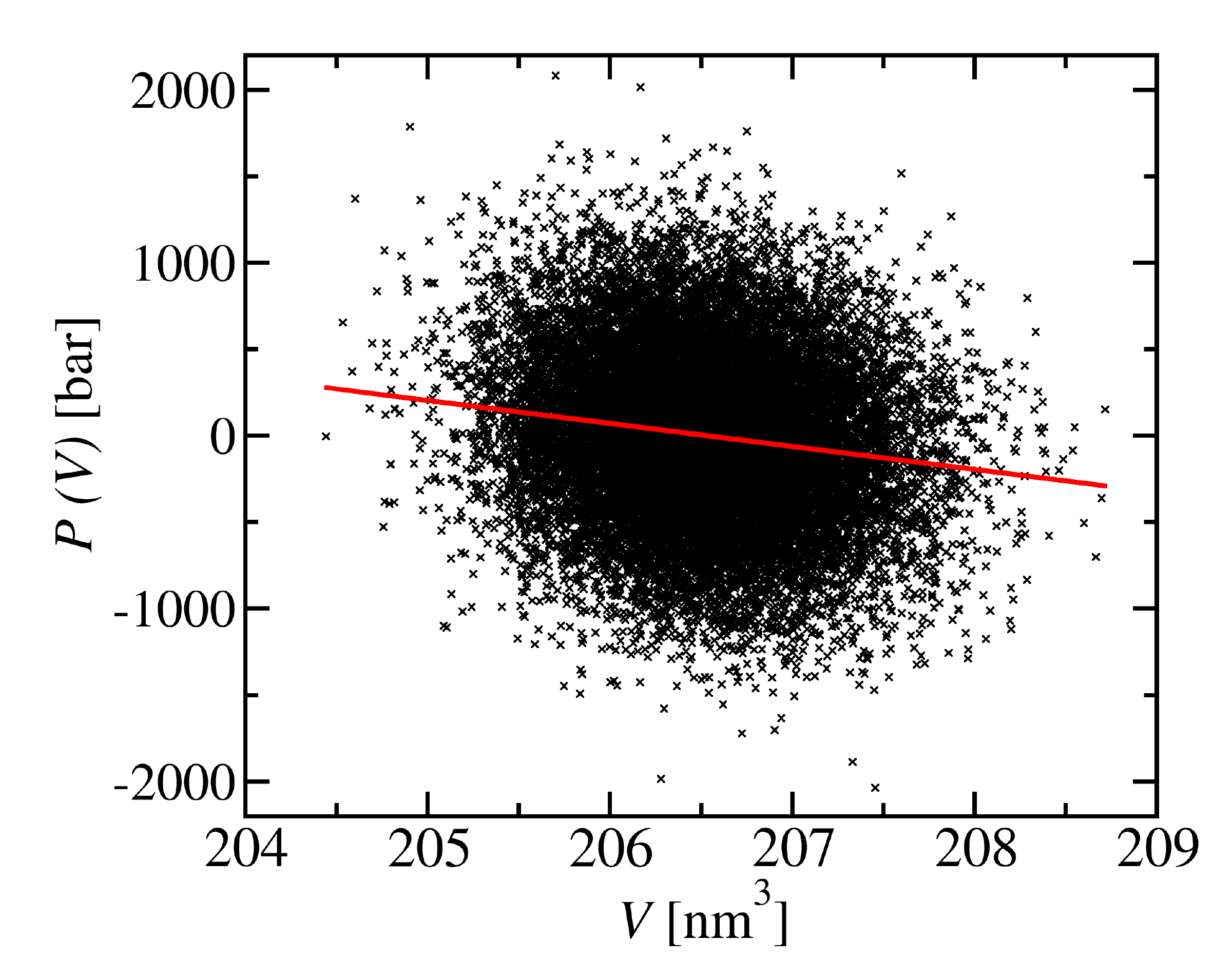}
  \caption{Accessed volume and corresponding $z$-pressure values, sampled from an $NPT$ simulation of the aspirin (001)/water interface. The linear fit to obtain the inverse compressibility is displayed by the straight line.}
  \label{fig:P_over_V}
\end{figure}

Our estimate of the isothermal compressibility of the aspirin/water interface is based on the associated volume/pressure fluctuations, as obtained in a long equilibrium simulation at constant pressure along the direction of the surface normal. 
Plotting the accessed pressure values over the corresponding cell volumes (cf. Fig. \ref{fig:P_over_V}) we calculate the inverse compressibility $\kappa^{-1}$ by a linear fit, according to 
\begin{equation}
  \frac{dP}{dV}=-\frac{1}{\kappa V} \, \quad .
\end{equation}
In spite of the pronounced scattering of the values, we obtain a meaningful slope with an error of 5 \%. We thus arrive at an interface compressibility of $\kappa_{\rm inter} = 3.6 \pm 0.2 \cdot 10^{-5}$ bar$^{-1}$, which is similar to the compressibility of pure water, $\kappa_{\rm water} = 4.5 \cdot 10^{-5}$ bar$^{-1}$.

%

\end{document}